# Acceleration Measurement Enhances the Bandwidth of Disturbance Observer in Motion Control Systems

Emre Sariyildiz
School of Mechanical, Materials, Mechatronic and Biomedical Engineering,
University of Wollongong, NSW Australia.
emre@uow.edu.au

***Abstract*- The trade-off between the noise-sensitivity and the performance of disturbance estimation is well-known in the Disturbance Observer- (DOb-) based motion control systems. As the bandwidth of the DOb increases, not only the performance but also the frequency range of disturbance estimation improves yet the motion controller becomes more sensitive to the noise of measurement system. This trade-off is generally explained by considering only the noise of sensors such as encoders. However, the digital implementation of the robust motion controller may significantly influence the noise sensitivity and performance of disturbance estimation in practice. This paper shows that the conventional DOb implemented by estimating velocity is subject to *waterbed effect* when the design parameters (i.e., sampling-time, nominal plant parameters and the bandwidth of the DOb) are not properly tuned in the digital motion controller synthesis. Therefore, the bandwidth of disturbance estimation is limited by *waterbed effect* in addition to the noise of velocity measurement system. To facilitate the digital motion controller synthesis, the design constraints of the conventional DOb are analytically derived in this paper. When the digital motion controller is implemented by estimating acceleration, *waterbed effect* does not occur, and good robust stability and performance can be achieved for all values of the design parameters of the acceleration measurement-based DOb. The bandwidth of disturbance estimation, however, cannot be freely increased due to the noise of acceleration sensors in practice. By employing Bode Integral Theorem in the discrete-time domain, the design constraints of the DOb-based digital motion control systems are clearly explained and it is shown that acceleration measurement can be used to enhance the bandwidth of the DOb, i.e., the performance and frequency range of disturbance estimation. To verify the proposed analysis and synthesis methods, simulation results are given for the DOb-based position and force control systems.***

## I. INTRODUCTION

The DOb is one of the most widely used motion control tools that is employed to estimate the internal and external disturbances (e.g., dynamic model uncertainties, friction, backlash, and contact force in physical interaction) of a servosystem [1–4]. It has been used to perform robust motion control and sensorless force control in various engineering systems, spaning from automobiles, hard-disks, satellites and unmanned aerial vehicles to surgery, rehabilitation and soft robots [2, 5–11]. It is a well-known fact that the performance of disturbance estimation is highly related to the bandwidth of the DOb, i.e., the higher the bandwidth of the DOb is the more the performance of disturbance estimation improves [12, 13]. However, the bandwidth of the DOb cannot be freely increased due to theoretical and practical limitations in real-world applications [12, 13].

In the DOb-based robust motion control technique, the disturbances of a servosystem are cancelled by feedbacking the reverse of the estimated disturbance signal in an inner-loop. This allows one to synthesise a performance controller (e.g., a Proportional-Derivative (PD) controller for position control and a Proportional (P) controller for force control) by considering only the nominal plant model of a servosystem in an outer-loop [12, 13]. Since the robustness and performance of the motion controller are independently adjusted in the inner- and outer- loop, respectively, the DOb-based robust motion controller has 2-Degrees-of-Freedom (DoF) [2]. As the bandwidth of the DOb increases, the robustness of the motion controller improves by increasing the suppression of disturbances in a wider frequency range in the inner-loop. This also allows one to increase the bandwidth of the outer-loop controller, and thus the performance of the motion control system. However, increasing the bandwidth of the DOb degrades not only the noise-sensitivity but also the stability of the digital motion control systems [14–16].

In the DOb-based robust sensorless force control technique, the contact force is estimated by identifying the internal disturbances of a servosystem. Similarly, the robustness is achieved in the inner-loop and sensorless force control is performed by feedbacking the estimated contact force in the outer-loop [17]. The DOb-based sensorless force estimation technique has several superiorities over force sensors [17–21]. For example: i) the bandwidth of the DOb is higher than that of a force sensor, so the interaction force is accurately detected within a larger frequency range and the stability of contact motion is improved [19], ii) the DOb explicitly detects contact force, so it does not introduce mechanical compliance to a servosystem [21], and iii) the cost and design complexity can be significantly reduced with the force sensorless force estimation technique [17, 21]. The performance of disturbance estimation is, however, directly related to the identification of internal disturbances and the dynamics of the DOb.

To improve the stability and performance of the DOb-based robust motion control systems, several analysis and synthesis methods have been proposed in the literature. Sariyildiz and Ohnishi derived the design constraints of the DOb in the continuous-time domain [21]. Murakami et al. and Ugurlu et al. designed model-based force controllers to improve the performance of contact force estimation [17, 22]. Sariyildiz and Ohnishi proposed an adaptive controller to automatically tune the robust stability and performance of the DOb-based force control systems [23]. Tsuji et al. showed that the trade-off between the noise sensitivity and performance of disturbance estimation changes when the DOb is synthesised by using different measurement methods [24]. Katsura et al. combined acceleration and position measurements to improve the performance of disturbance force estimation by enhancing the bandwidth of the DOb [25]. Mitsantisuk et al. and Antonello et al. used Kalman filter to suppress the noise of acceleration measurement in the DOb synthesis [26, 27]. Nevertheless, the design constraints of the DOb-based digital motion control systems have not been analytically derived yet. Moreover, the effect of velocity and acceleration measurements on the design constraints of the DOb (e.g., the upper bound of the bandwidth of disturbance estimation) has yet to be clearly explained in the literature.

In this paper, the DOb-based motion control systems are analysed by employing Bode Integral Theorem in discrete-time [28, 29]. It is shown that the robust stability and performance of the digital motion controller significantly change when velocity and acceleration measurements are employed in the DOb synthesis. The velocity measurement-based DOb (i.e., the conventional DOb) is subject to *waterbed effect* when the nominal plant parameters and the bandwidth of the observer are not properly tuned. As the performance of disturbance estimation is improved by increasing the bandwidth of the DOb, the peaks of the sensitivity and complementary sensitivity functions increase at middle/high frequencies. This makes the DOb more sensitive to noise and degrades its robust stability and performance. In addition to the trade-off between the noise sensitivity of the velocity measurement system and the performance of force estimation, the bandwidth of the conventional DOb is limited by *waterbed effect*. To relax the design constraints on the bandwidth of the observer, acceleration measurement-based DOb is proposed in this paper. It is shown that *waterbed effect* does not occur when the bandwidth of the acceleration measurement-based DOb is increased to improve the robustness of the motion control system. Good robust stability and performance can be achieved for all values of the bandwidth of the acceleration measurement-based DOb. However, the bandwidth of the DOb is limited by the noise sensitivity of the acceleration measurement system, so the performance of disturbance estimation cannot be freely improved in practice. In other words, the trade-off between the noise sensitivity of the accelerometer and the performance of force estimation is the only design constraint on the bandwidth of the acceleration measurement-based DOb. This paper analytically derives the design constraints of the DOb in the discrete-time domain. The proposed design constraints allow one to systematically synthesise high-performance digital motion control systems. Simulation results are given to verify the proposed analysis and synthesis methods.

It is noted that continuous-time analysis methods fall-short in deriving the robust stability and performance constraints of the DOb implemented by computers and microcontrollers. Therefore, unexpected dynamic responses, such as poor stability and performance, may be observed when the design parameters of the digital motion controller (e.g., the bandwidth of the DOb) are tuned in the continuous-time domain. For the sake of brevity, this paper conducts analysis only in the discrete-time domain. The reader is recommended to refer to [2, 13, 30] for further details on the continuous-time analysis and synthesis of the DOb-based motion control systems.

The rest of the paper is organised as follows. The conventional DOb-based digital robust motion controller, which is synthesised by employing velocity measurement, is presented in section II. To improve the robust stability and performance of the digital motion controller, the acceleration measurement-based DOb is proposed in section III. DOb-based force-sensorless force controller is presented in section IV. Simulation results and concluding remarks are given in section V and section VI, respectively.

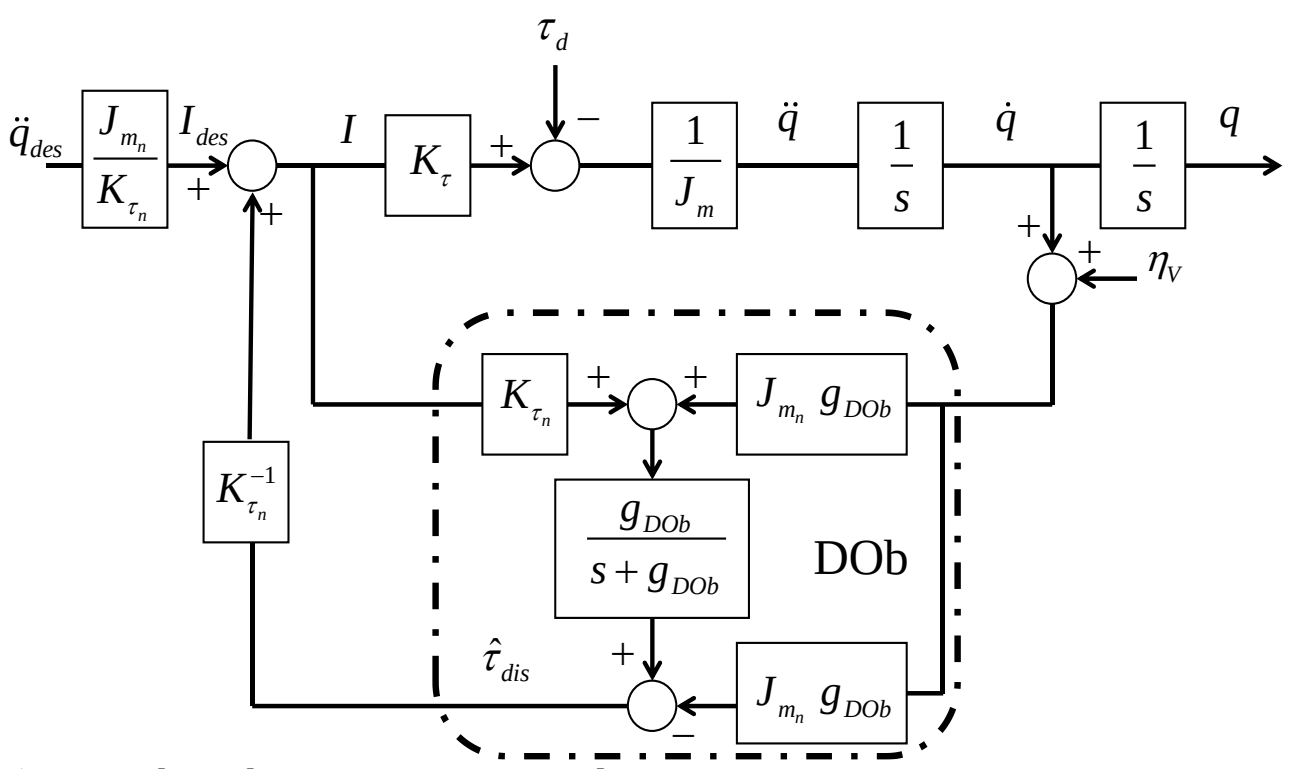

a) DOb in the continuous-time domain.

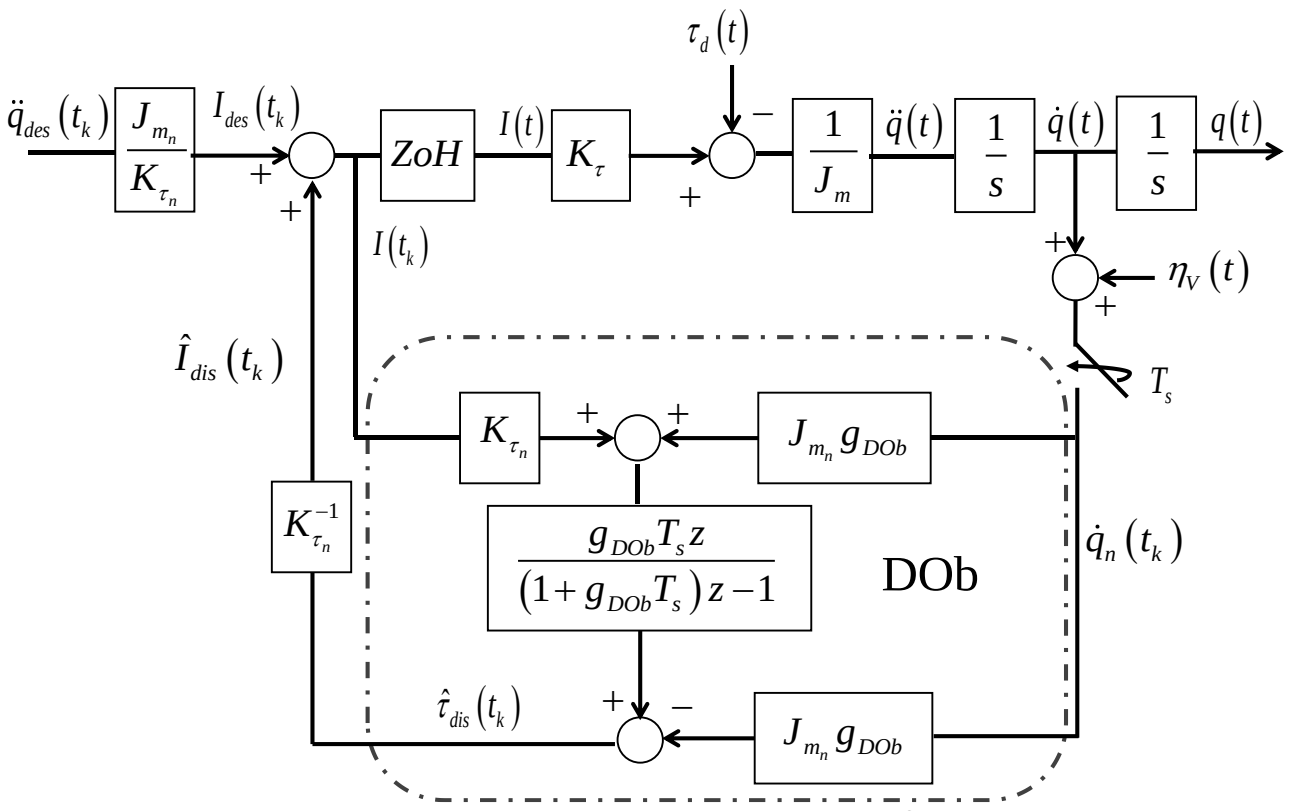

b) DOb in the discrete-time domain. Low-pass-filter is synthesised by employing Backward Euler integration method.

Fig. 1: Block diagrams of the conventional velocity measurement-based DOb in the continuous- and discrete- time domains.

## II. Disturbance Observer

Figure 1 illustrates the conventional DOb-based motion controller in the continuous- and discrete- time domains [16, 21]. In this figure, the following apply:

| | |
|---|---|
| $J_m$ and $J_{m_n}$ | uncertain and nominal inertiae; |
| $K_\tau$ and $K_{\tau_n}$ | uncertain and nominal thrust coefficients; |
| $\tau_d$ and $\eta_V$ | disturbance and noise exogenous inputs; |
| $q$, $\dot{q}$ and $\ddot{q}$ | angle, velocity and acceleration; |
| $g_{DOb}$ | bandwidth of the DOb; |
| $I$ | current of a DC motor; |
| $\tau_{dis}$ and $I_{dis}$ | disturbance torque and current; |
| $ZoH$ | Zero order Hold; |
| $T_s$ | Sampling Time; |
| $t, t_k = kT_s$ | Time in continuous and discrete domains; |
| $s$, $z = e^{sT_s}$ | complex variables; |
| $\hat{\bullet}$ | estimation of $\bullet$; |
| $\bullet_{des}$ | desired $\bullet$. |

The dynamic equations of the servosystem and disturbance estimation are derived from this figure as follows.

In the continuous-time domain:

$$\dot{q}(s) = J_m^{-1} K_\tau \frac{I(s)}{s} - J_m^{-1} \frac{\tau_d(s)}{s} \tag{1}$$

$$\begin{aligned}\hat{\tau}_{dis}(s) &= \left(K_{\tau_n} I(s) + J_{m_n} g_{DOb} \dot{q}_n(s)\right) \frac{g_{DOb}}{s + g_{DOb}} - J_{m_n} g_{DOb} \dot{q}_n(s) \\ &= \left(K_{\tau_n} I(s) - J_{m_n} \ddot{q}_n(s)\right) \frac{g_{DOb}}{s + g_{DOb}} = \tau_{dis}(s) \frac{g_{DOb}}{s + g_{DOb}}\end{aligned} \tag{2}$$

where $\ddot{q}_n(s) = s\dot{q}_n(s)$ and $\tau_{dis}(s) = \tau_d(s) + \Delta K_\tau I + \Delta J \ddot{q}_n$ in which $\Delta K_\tau$ and $\Delta J$ represent parametric uncertainties.

In the discrete-time domain:

$$\dot{q}(z) = J_m^{-1} K_\tau \frac{T_s}{z-1} I(z) - J_m^{-1} \frac{T_s}{z-1} \tau_d(z) \quad (3)$$

$$\begin{aligned}
\hat{\tau}_{dis}(z) &= \left(K_{\tau_n} I(z) + J_{m_n} g_{DOb} \dot{q}_n(z)\right) \frac{g_{DOb} T_s z}{(1+g_{DOb}T_s)z-1} - J_{m_n} g_{DOb} \dot{q}_n(z) \\
&= \left(K_{\tau_n} I(z) - \frac{z-1}{T_s z} J_{m_n} \dot{q}_n(z)\right) \frac{g_{DOb} T_s z}{(1+g_{DOb}T_s)z-1} \quad (4) \\
&= \left(K_{\tau_n} I(z) - J_{m_n} \ddot{q}_n(z)\right) \frac{g_{DOb} T_s z}{(1+g_{DOb}T_s)z-1} = \tau_{dis}(z) \frac{g_{DOb} T_s z}{(1+g_{DOb}T_s)z-1}
\end{aligned}$$

where $\dot{q}_n(z) = \frac{T_s z}{z-1} \ddot{q}_n(z)$ when Backward Euler method is employed.

The transfer functions of the conventional DOb-based digital robust motion controller can be derived by using Fig. 1b and Eqs. (3) and (4) as follows.

$$\ddot{q}(z) = \alpha \frac{(1+g_{DOb}T_s)z-1}{z-(1-\alpha g_{DOb}T_s)} \ddot{q}_{des}(z) - \frac{1}{J_m} \mathrm{S_V}(z) \tau_d(z) - \frac{(z-1)}{T_s} \mathrm{T_V}(z) \eta_V(z) \quad (5)$$

where $\mathrm{S_V}(z) = \frac{z-1}{z-(1-\alpha g_{DOb}T_s)}$ and $\mathrm{T_V}(z) = \frac{\alpha g_{DOb} T_s}{z-(1-\alpha g_{DOb}T_s)}$ are the sensitivity and complementary sensitivity transfer functions in which $\alpha = \left(J_{m_n} K_\tau\right) / \left(J_m K_{\tau_n}\right)$.

The stability and performance of the digital motion controller can be adjusted by tuning the phase-lead/lag compensator given in Eq. (5). As $\alpha$ is increased (i.e., the nominal inertia/thrust coefficient is increased/decreased) the phase margin of the digital motion controller improves. However, the stability constraint puts an upper bound on the design parameter $\alpha$, because the conventional DOb-based digital motion controller becomes unstable when $\alpha g_{DOb} T_s > 2$ and exhibits oscillatory response when $\alpha g_{DOb} T_s > 1$. Eq. (5) also shows that as the bandwidth of the DOb increases, the sensitivity function gets smaller values at low frequencies, i.e., the robustness against disturbances improves. However, not only the phase margin but also the bandwidth of the DOb is limited by the stability constraint given by $\alpha g_{DOb} T_s < 1$. To improve the phase margin and the robustness against disturbances, the sampling time of the digital motion controller should be decreased, but this generally increases the cost in engineering applications.

Let us now consider the robustness of the conventional DOb by employing Bode Integral Theorem in the discrete-time domain. The Bode integral equation of the DOb shown in Fig. 1b as follows.

$$\int_{-\pi}^{\pi} \ln\left|\mathrm{S_V}\left(e^{j\omega T_s}\right)\right| d\omega T_s = -2\pi \ln\left|1 + \lim_{z\to\infty} \mathrm{L_V}(z)\right| = 0 \quad (6)$$

where $\mathrm{L_V}(z) = \frac{\alpha g_{DOb} T_s}{z-1}$ is the open-loop transfer function of the conventional velocity measurement-based DOb [29].

Equation (6) shows that as the robustness against disturbances is improved by decreasing $\left|\mathrm{S_V}\left(e^{j\omega T_s}\right)\right|$ at low frequencies, the peak of the sensitivity function increases at middle/high frequencies to hold the Bode's integral equation. In other words, the conventional DOb-based digital motion controller is subject to *waterbed effect* when the design parameters are not properly tuned. For example, as the bandwidth of the DOb is increased to improve the performance of force estimation, the robust stability and performance of the motion controller deteriorate due to the high-peak values of the sensitivity and complementary sensitivity transfer functions.

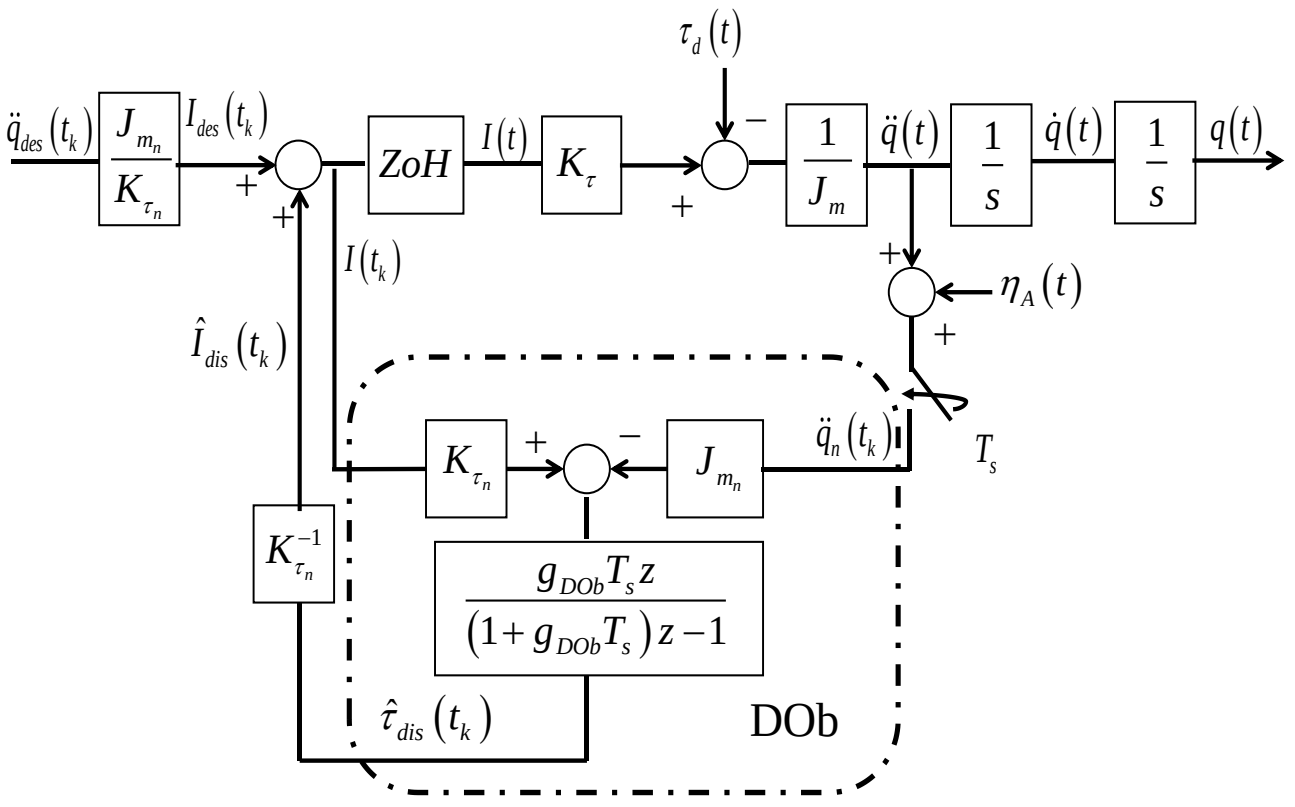

a) Acceleration measurement-based DOb.

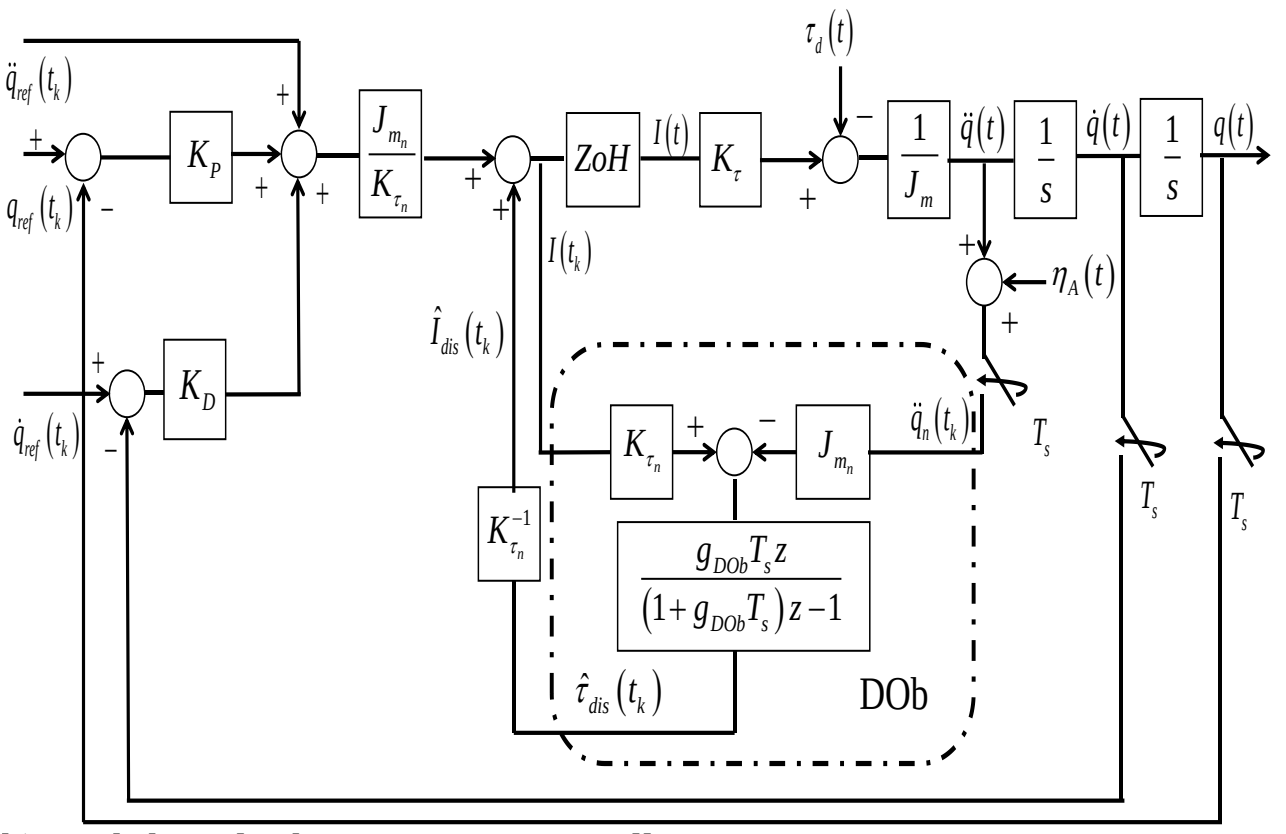

b) DOb-based robust position controller.

Fig. 2: Block diagrams of the acceleration measurement-based DOb and the robust position controller in the discrete-time domain.

To achieve good robust stability and performance, the constraints on the design parameters of the DOb can be derived by using the frequency responses of the sensitivity and complementary sensitivity functions. For example, Eqs. (7) and (8) should hold to achieve $\left|\mathrm{S_V}(z)\right|_{\max} \le 1/\Gamma_{\mathrm{S_V}}$ and $\left|\mathrm{T_V}(z)\right|_{\max} \le 1/\Gamma_{\mathrm{T_V}}$.

$$\alpha g_{DOb} T_s \le 2\left(1-\Gamma_{\mathrm{S_V}}\right) \quad (7)$$

$$\alpha g_{DOb} T_s \le 2/\left(1+\Gamma_{\mathrm{T_V}}\right) \quad (8)$$

where $0 < \Gamma_{\mathrm{S_V}} < 1$ and $0 < \Gamma_{\mathrm{T_V}} < 1$ [16].

## III. Acceleration Measurement-Based DOb: DOb-Based Robust Position Control System

Figure 2 illustrates the block diagrams of the acceleration measurement-based DOb and the robust position controller in the discrete-time domain. In this figure, $\eta_A$ represents the noise of acceleration sensor, and $K_P$ and $K_D$ represent the PD control gains. The other parameters are described earlier.

The dynamic equations of the servosystem and disturbance estimation are derived from Fig. 2a as follows.

$$\ddot{q}(t_k) = J_m^{-1} K_\tau I(t_k) - J_m^{-1}\tau_d(t_k) \tag{9}$$

$$\hat{\tau}_{dis}(z) = \left(K_{\tau_n} I(z) - J_{m_n}\ddot{q}_n(z)\right)\frac{g_{DOb}T_s z}{(1+g_{DOb}T_s)z-1} \tag{10}$$

where $\ddot{q}_n(t_k) = \ddot{q}(t_k) + \eta_A(t_k)$ is the measured acceleration state that includes the noise of acceleration sensor.

By substituting Eq. (10) into Eq. (9), the transfer functions of the digital motion controller illustrated in Fig. 2a can be derived as follows.

$$\ddot{q}(z) = \alpha\frac{(1+g_{DOb}T_s)z-1}{(1+\alpha g_{DOb}T_s)z-1}\ddot{q}_{des}(z) - \frac{1}{J_m}\mathrm{S_A}(z)\tau_d(z) - \mathrm{T_A}(z)\eta_A(z) \tag{11}$$

where $\mathrm{S_A}(z) = \frac{z-1}{(1+\alpha g_{DOb}T_s)z-1}$ and $\mathrm{T_A}(z) = \frac{\alpha g_{DOb}T_s z}{(1+\alpha g_{DOb}T_s)z-1}$ are the sensitivity and complementary sensitivity transfer functions of the acceleration measurement-based DOb.

Similar to the velocity measurement-based DOb, Eq. (11) shows that the stability and performance of the acceleration measurement-based DOb can be adjusted by tuning the phase-lead/lag compensator. A phase-lead (phase-lag) compensator is obtained when $\alpha > 1$ $(\alpha < 1)$, and the phase-margin of the digital motion controller improves as $\alpha$ is increased. Moreover, the sensitivity function $\mathrm{S_A}(z)$ gets smaller values, i.e., the robustness against disturbances improves, at low frequencies as the bandwidth of the DOb increases. The acceleration measurement-based DOb is, however, stable for all values of the design parameters of $\alpha$ and $g_{DOb}$. Therefore, the phase margin and the bandwidth of the DOb (i.e., the performance of disturbance estimation) can be freely increased without degrading the stability of the digital motion controller.

Let us now consider the robustness of the acceleration measurement-based DOb. The Bode integral equation of the DOb shown in Fig. 2a is as follows.

$$\begin{aligned}\int_{-\pi}^{\pi}\ln\left|\mathrm{S_A}\left(e^{j\omega T_s}\right)\right|d\omega T_s &= -2\pi\ln\left|1+\lim_{z\to\infty}\mathrm{L_A}(z)\right| \\ &= -2\pi\ln\left|1+\alpha g_{DOb}T_s\right|\end{aligned} \tag{12}$$

where $\mathrm{L_A}(z) = \alpha g_{DOb}T_s\frac{z}{z-1}$ is the open-loop transfer function of the acceleration measurement-based DOb [16, 29].

As the phase-margin and the bandwidth of the DOb (i.e., $\alpha$ and $g_{DOb}$) are increased, the right-hand side of Eq. (12) gets smaller values. Therefore, the Bode's integral equation can hold without exhibiting a high-sensitivity peak as the robustness of the digital motion controller is improved by decreasing the sensitivity function at low frequencies. Since the acceleration measurement-based DOb is not subject to *waterbed effect*, the bandwidth and accuracy of disturbance estimation can be freely increased without degrading the robust stability and performance of the digital motion controller.

Compared to the conventional DOb illustrated in Fig. 1, the design parameters of the acceleration measurement based DOb (i.e., $\alpha$ and $g_{DOb}$) are not constrained by *waterbed effect* and the stability of the digital motion controller. However, the noise of acceleration sensor limits the performance of force estimation, i.e., the bandwidth of the acceleration measurement-based DOb, in practice [21, 26, 27].

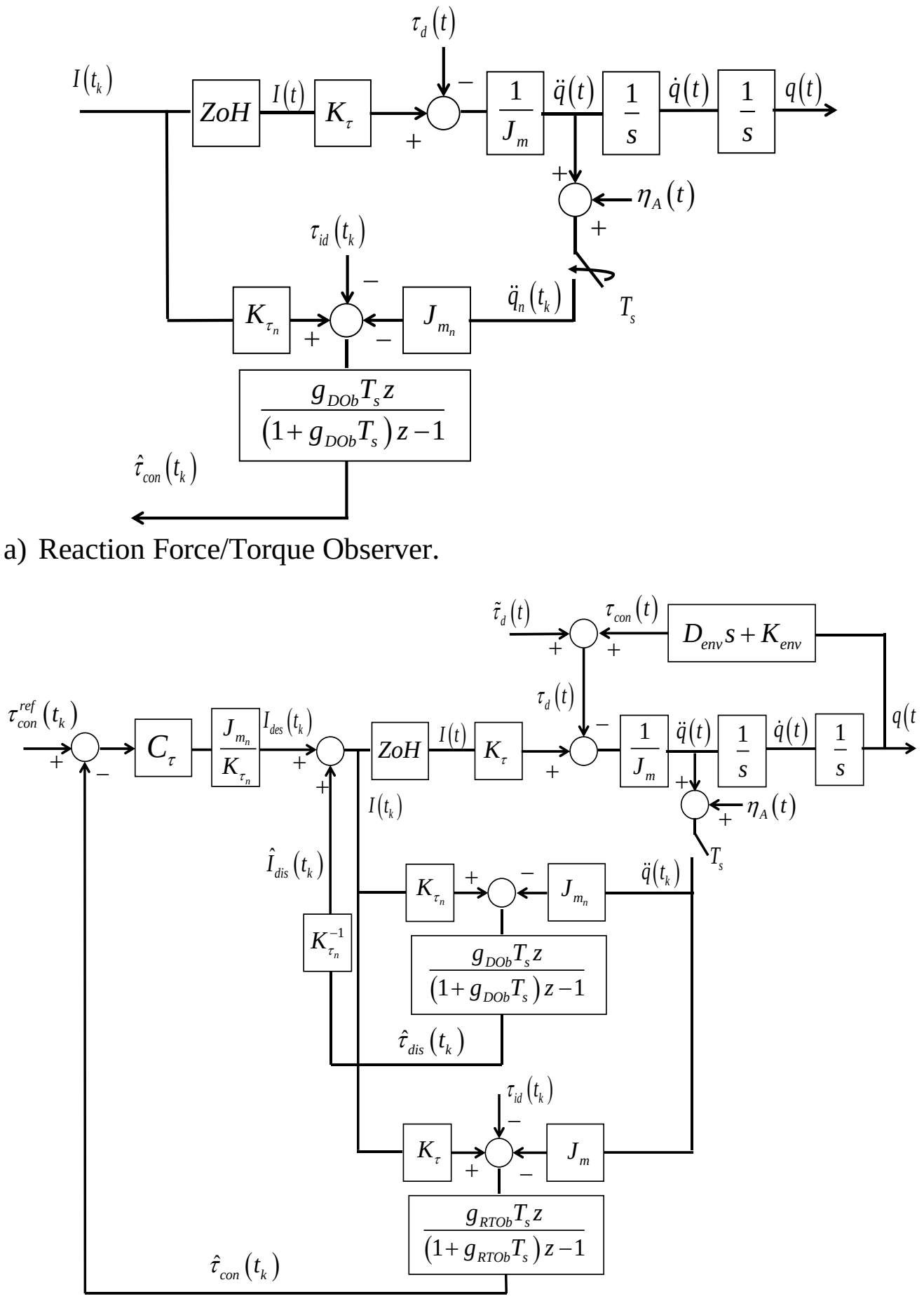

a) Reaction Force/Torque Observer.

b) DOb-based force-sensorless digital robust force controller.

Fig. 3: Block diagrams of the DOb-based digital force control systems.

## IV. DOb-based Robust Force Control System

Equations (2), (4) and (10) show that internal and external disturbances (e.g., interactive dynamic forces of a multibody system such as gravitational and Coriolis forces, internal disturbances due to uncertainties, friction, hysteresis and backlash, and external disturbances due to contact motion and load) are lumped together into a fictitious variable $\tau_{dis}$ and this variable is estimated by the DOb. To estimate the contact force, disturbances acting on the system should be identified and feedforwarded to the DOb as illustrated in Fig. 3. This is called Reaction Force/Torque Observer in the literature [17]. In this figure, $\tau_{id}$ is the identified disturbance, and $\tau_{con}$ and $\hat{\tau}_{con}$ are the contact force and its estimation, respectively.

By using Fig. 3a and a simple mathematical manipulation, the dynamic equations of the contact force estimation are derived as follows.

$$J_{m_n}\ddot{q}(t_k) = K_{\tau_n} I(t_k) - \tau_{con}(t_k) - \tilde{\tau}_{dis}(t_k) \tag{13}$$

$$\begin{aligned}\hat{\tau}_{con}(z) &= \left(K_{\tau_n} I(z) + J_{m_n} g_{DOb}\dot{q}_n(z) - \tau_{id}(z)\right)\frac{g_{DOb}T_s z}{(1+g_{DOb}T_s)z-1} - J_{m_n} g_{DOb}\dot{q}_n(z) \\ &= \tau_{con}(z)\frac{g_{DOb}T_s z}{(1+g_{DOb}T_s)z-1} - \left(\tau_{id}(z) - \tilde{\tau}_{dis}(z)\right)\frac{g_{DOb}T_s z}{(1+g_{DOb}T_s)z-1}\end{aligned} \tag{14}$$

where $\tilde{\tau}_{dis}(t_k) = \tau_{dis}(t_k) - \tau_{con}(t_k)$.

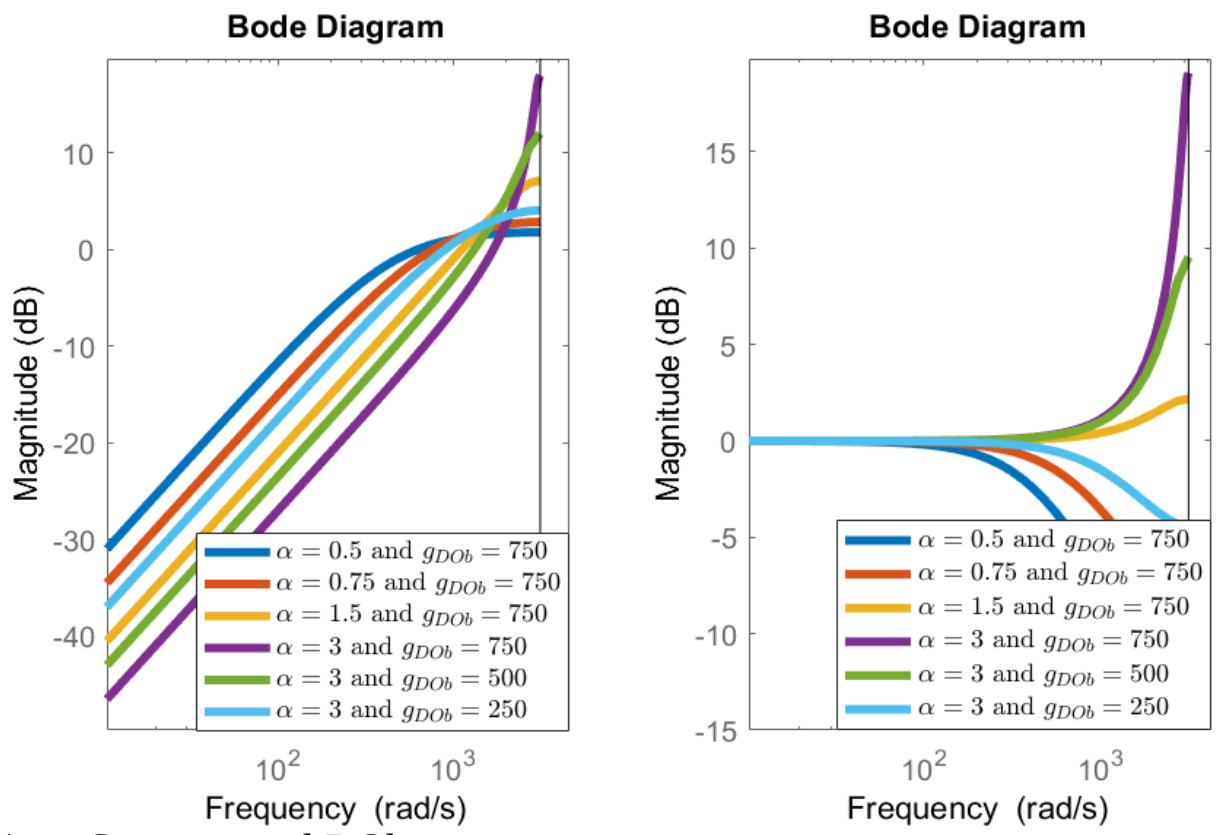

a) Conventional DOb.

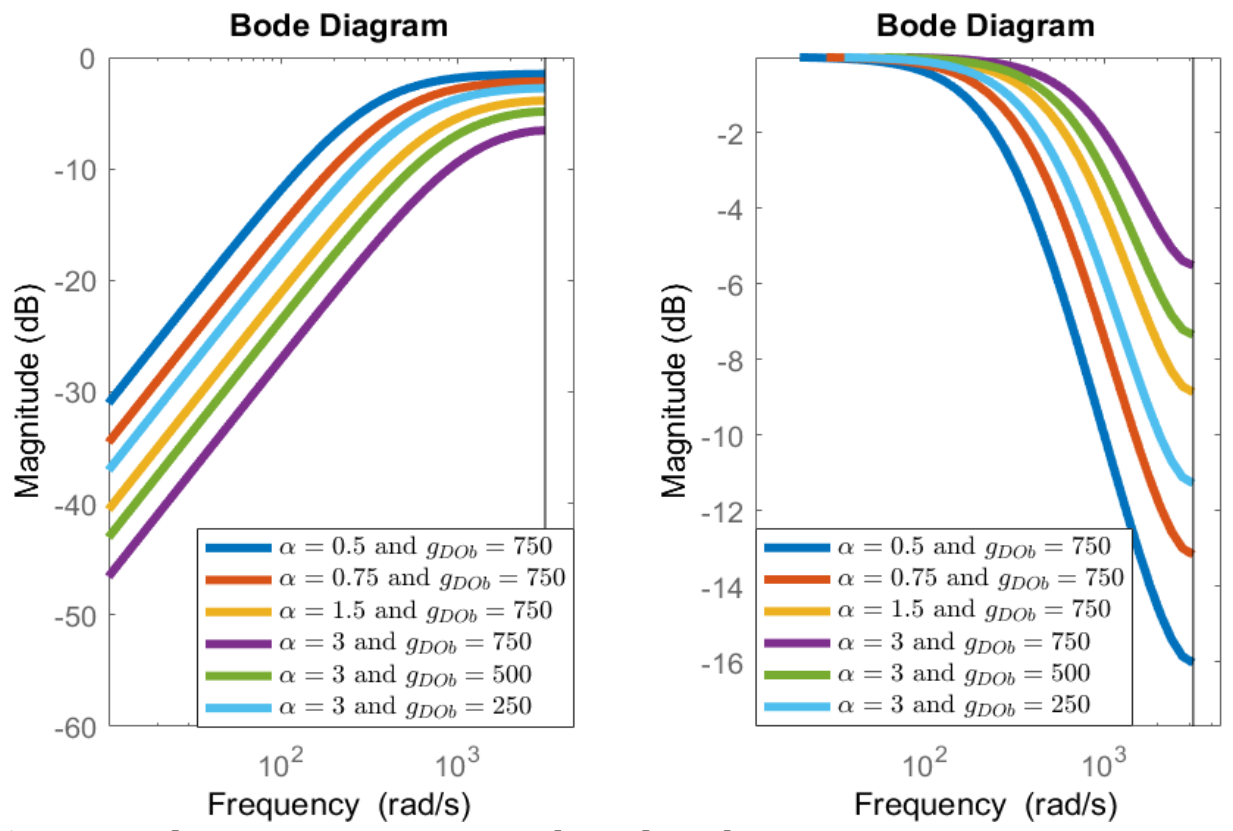

b) Acceleration measurement-based DOb.

Fig. 4: Sensitivity (left-figures) and complementary sensitivity (right figures) functions' frequency responses.

Equations (13) and (14) show that the performance of contact force estimation is directly influenced by the bandwidth of the DOb and the identification of system disturbances [17, 21, 23]. To accurately estimate the contact force, the feedforward control input of the DOb should be tuned by using $\tau_{id}(t_k) = \tilde{\tau}_{dis}(t_k)$. Compared to the force sensor, the main drawback of the DOb-based force-sensorless force control technique is that it requires the dynamic model of the motion control system to accurately estimate the contact force. The reader is referred to [21] for a model-based force controller synthesis and to [23] for an adaptive robust force controller synthesis.

Figure 3b illustrates the DOb-based force-sensorless digital robust force controller that is synthesised by combining two DObs in the inner- and outer- loops. In this figure, $g_{RTOb}$ represents the bandwidth of the DOb in the outer-loop. While the robustness of the system is improved by employing the DOb in the inner-loop, the performance of the force controller is adjusted in the outer-loop. For the sake of brevity, the robust force controller is not considered in this paper. The reader is referred to [21, 23] for further details on the DOb-based robust force controller analysis and synthesis.

## V. Simulation Results

In this section, simulation results are given to verify the proposed analysis and synthesis methods. The parameters of the simulations are $J_m = 0.01$, $K_\tau = 0.25$, $C_\tau = 150$, $D_{env} = 1$, $K_P = 2500$, $K_D = 125$ and $T_s = 1$ ms.

Let us start by considering the robust stability and performance of the DOb. Figure 4 illustrates the frequency responses of the sensitivity and complementary sensitivity transfer functions given in Eqs. (5) and (11). It is clear from this figure that the robustness against disturbances improves at low frequencies as $\alpha$ and $g_{DOb}$ are increased. This increases the peaks of the sensitivity and complementary sensitivity functions (i.e., *waterbed effect* occurs) when the DOb is implemented by using velocity measurement as shown in Fig. 4a. However, good robust stability and performance can be achieved for all values of the design parameters of $\alpha$ and $g_{DOb}$ in the acceleration measurement-based DOb (Fig. 4b).

Let us now consider the stability of the DOb-based robust position control system shown in Fig. 2b. Figure 5 illustrates the root-loci of the robust position controller when the DOb is synthesised by employing velocity and acceleration measurements. The bandwidth of the observer is set at 750 rad/s. This figure shows that the phase-lag compensator (i.e., $\alpha < 1$) degrades the stability of the robust position controller. Also, the stability deteriorates when the design parameters of the conventional DOb are not properly tuned. However, the stability improves as $\alpha$ is increased when the acceleration measurement-based DOb is employed in the robust position controller synthesis.

Last, let us consider the stability and performance of the DOb-based digital robust force controller. Force regulation control responses of the digital force controller implemented by using conventional and acceleration measurement-based DObs are illustrated in Fig. 6. When the environment is relatively soft (e.g., $K_{env}$ is 100 Nm/rad in the simulations), both velocity and acceleration measurement-based digital robust force controllers provide high-performance force control (see black curves in Figs. 6a and 6b). The oscillation of the force control response increases, i.e., the stability of the digital robust force controllers degrades, when the stiffness of the environmental impedance increases. This is illustrated by using blue curves in Fig. 6. To tackle this problem, the phase-margin of the controller and the bandwidth of force estimation are improved by increasing $\alpha$ and $g_{RTOb}$. Although the stability of the digital robust force controller implemented by using the acceleration measurement-based DOb is improved, the conventional digital robust force controller becomes unstable as the bandwidth of force estimation increases. This is illustrated by using green curves in Fig. 6.

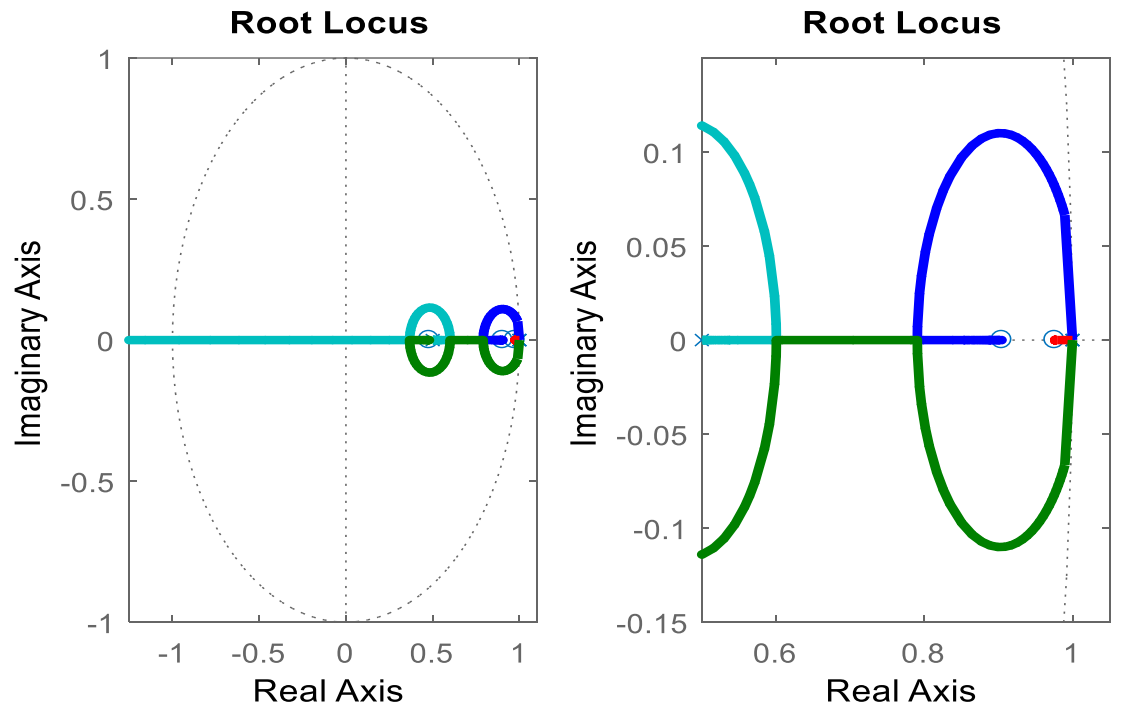

a) Conventional DOb.

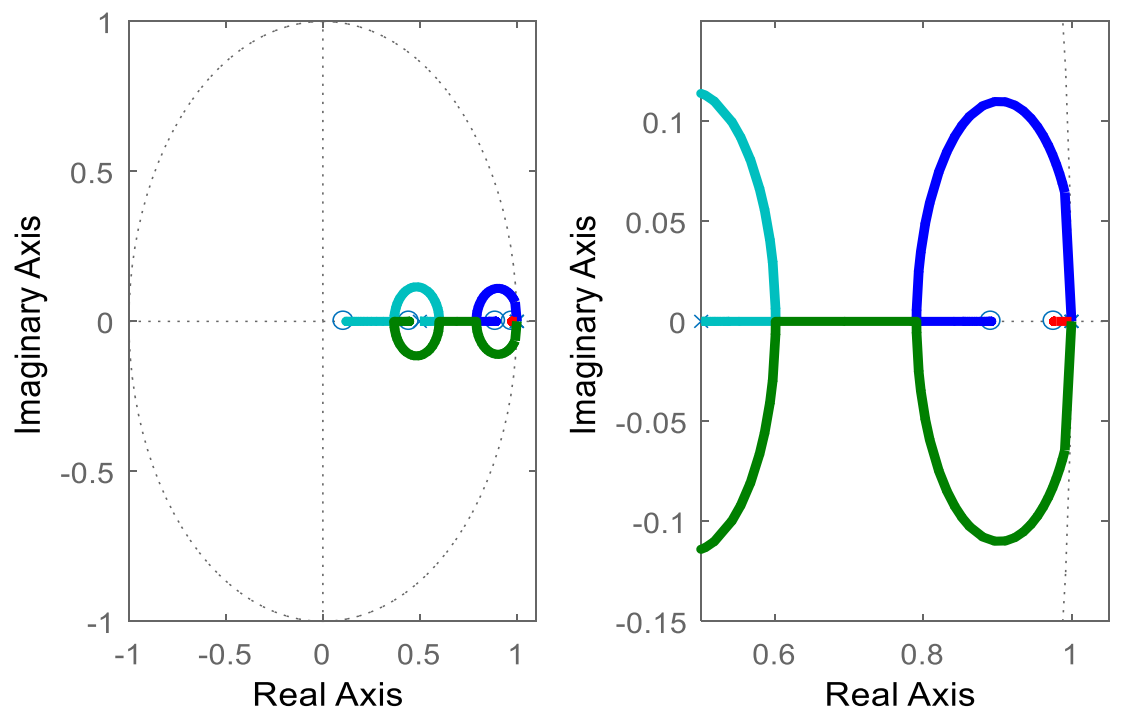

a) Acceleration measurement-based DOb.

Fig. 5: Root-loci of the robust position control system with respect to $\alpha$ .

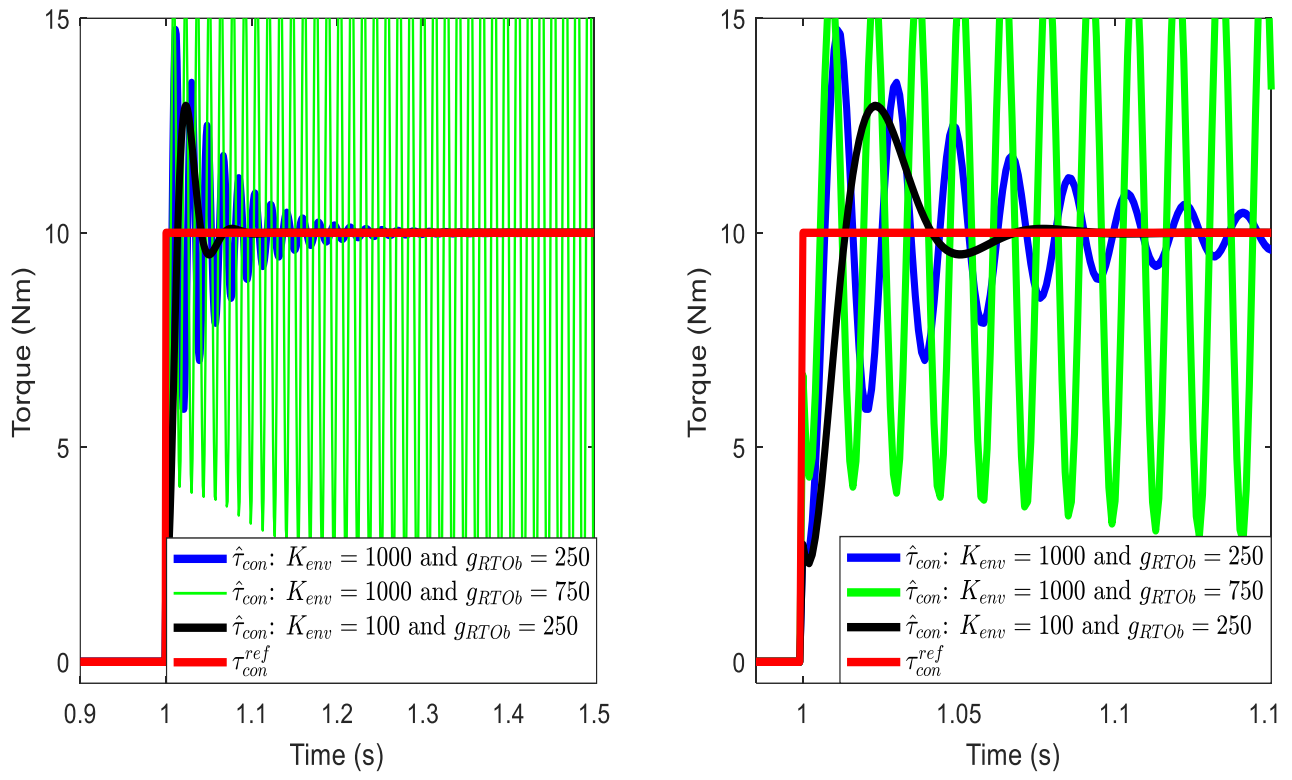

a) Conventional DOb.

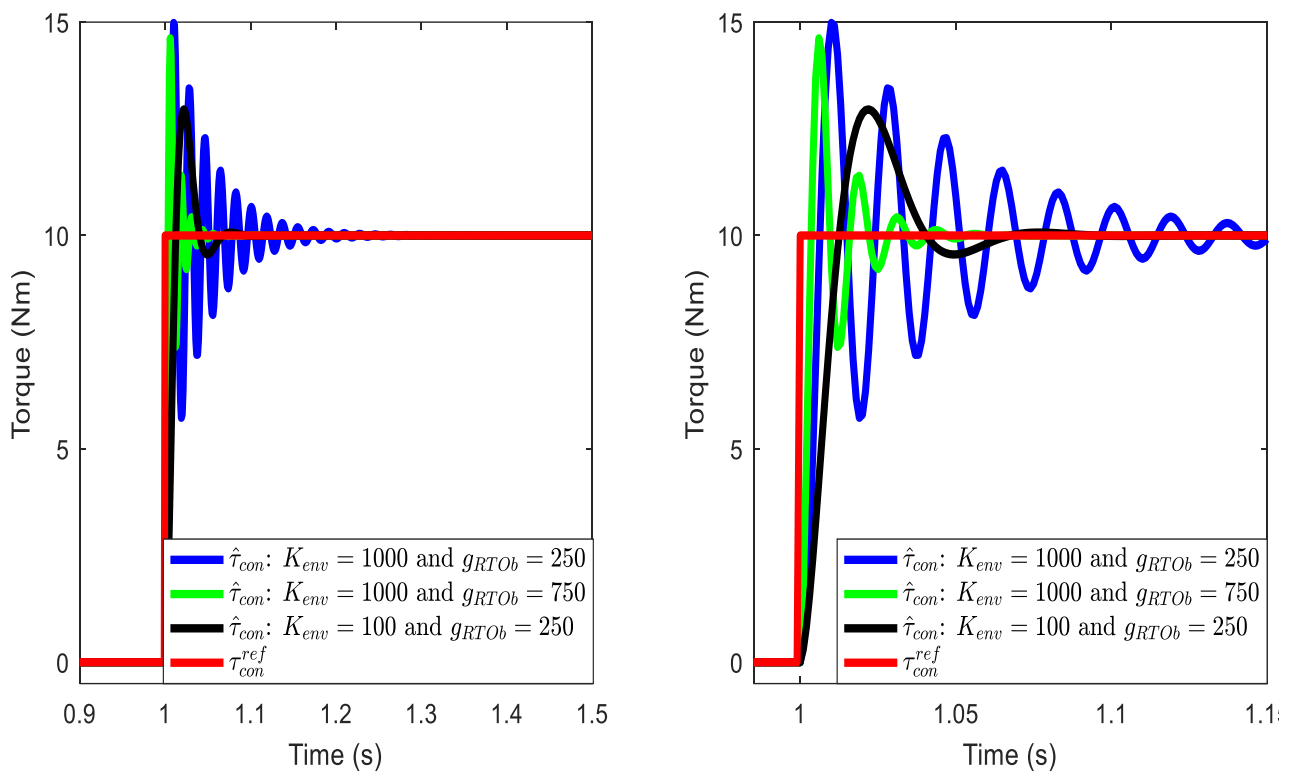

b) Acceleration measurement-based DOb.

Fig. 6: Force regulation control results of the digital robust force controller.

## VI. Conclusion

This paper shows that the design parameters of the conventional DOb-based digital motion controller (e.g., the nominal inertia and the bandwidth of the observer) are constrained not only by the noise of velocity measurement system but also by *waterbed effect*. For example, as the bandwidth of the DOb increases, the peaks of the sensitivity and complementary sensitivity functions become higher. This makes the digital motion controller more sensitive to the noise of velocity measurement system and degrades the robust stability and performance. When the DOb is implemented by using acceleration measurement, the digital motion controller is not subject to *waterbed effect*. The performance of force estimation can be improved without degrading the robust stability and performance. However, the bandwidth of the DOb is limited by the noise of acceleration measurement system in practice.